\documentclass[twocolumn,prl,showpacs,floatfix]{revtex4}
\usepackage{graphicx}
\usepackage{bm}

\begin{document}
\title{Towards unified understanding of conductance of stretched monatomic
contacts}
\author{H.-W. Lee,$^{1}$ H.-S. Sim,$^{2,3}$
D.-H. Kim,$^4$ and K. J. Chang$^4$}
%
%
\affiliation{$^1$Department of Physics, Pohang University of
Science and Technology, Pohang, Kyungbuk 790-784, Korea}
%
\affiliation{$^2$Max-Planck-Institut f\"{u}r Physik komplexer
Systeme, N\"{o}thnitzer Str. 38, 01187 Dresden, Germany}
%
\affiliation{$^3$Korea Institute for Advanced Study, 207-43
Cheongryangri-dong, Dongdaemun-gu, Seoul 130-722, Korea}
\affiliation{$^4$Department of Physics, Korea Advanced Institute
of Science and Technology, Taejon 305-701, Korea}
\date{\today}
\begin{abstract}
When monatomic contacts are stretched, their conductance behaves
in qualitatively different ways depending on their constituent
atomic elements.
Under a single assumption of resonance formation, we show that
various conductance behavior can be understood in a unified way
in terms of the response of the resonance to stretching.
This analysis clarifies the crucial roles played by the number of
valence electrons, charge neutrality, and orbital shapes.
\end{abstract}

\maketitle


Monatomic contacts with one-atom-wide cross-section (referred to
as ``contacts'' hereafter) have been realized by scanning
tunneling microscopes or mechanically controllable break
junctions~\cite{Ruitenbeek}. Upon stretching, the conductance $G$
of various contacts shows different behavior. In contacts made of
monovalent atoms such as Na and Au, only one channel contributes
to transport and $G$ stays at the quantized value $G_0 \equiv
2e^2/h$ during the
stretching.~\cite{Krans95Nature,Rubio,Ohnishi,Yanson,Rubio-Bollinger}.
In contacts made of polyvalent atoms such as Al and Pb, on the
other hand, current is carried by three channels and $G$ does not
stay at quantized values during the
stretching~\cite{Scheer,Yanson_comment,SanchezPortal,Cuevas_stretching}.
Thus monovalent and polyvalent contacts are qualitatively
different. Moreover there are differences even between polyvalent
contacts; during the stretching, $G$ increases (decreases) for Al
(Pb) contacts.

The different conductance features were addressed by numerical
calculations~\cite{Cuevas,Cuevas_stretching,Kobayashi,Yeyati,Sim,Hirose} and
some calculations reproduced experimental results for certain
contacts. However the calculational results are often too specific to the
particular contacts under calculation and 
the understanding on the origin of the aforementioned
difference is yet far from satisfactory.
For example, the connection between the conductance features and
the number of valence electrons is poorly understood. 
In this respect, it is desirable to have an alternative approach
that allows easier comparison of various contacts and 
thus provides an insightful explanation of the difference.
We also remark that numerical calculations of
conductance are sometimes subject to strong finite size
effects~\cite{Palacios}. 
Thus a naive comparison may be dangerous between experimental findings 
and numerical calculation results based on small cluster modelings of atomic
contacts.
For the behavior of Al contacts, existing
calculations~\cite{Cuevas_stretching,Kobayashi} have reported qualitatively
different results~\cite{difference} and we suspect that 
the difference may be related to the finite size effect.

In this paper, we present an analysis that allows one to capture
the connection between the conductance properties and microscopic
features such as the number of valence electrons, charge
neutrality, and orbital shapes. This connection provides a unified
understanding of various conductance behavior. This analysis is
applicable, provided that electron transport has resonant
character.
%


\begin{figure}
\includegraphics[width=0.40\textwidth]{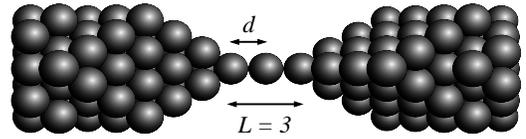}
\caption{ A model of a monatomic Al atom
contact~\protect\cite{L=1vsL=3}. } \label{Alstructure}
\end{figure}

The assumption of the resonance is very plausible in a stretched
contact. First of all, the central atom of a contact
(Fig.~\ref{Alstructure})~\cite{L=1vsL=3} is relatively weakly
coupled to its environment even without stretching since it has a
fewer number of nearest neighbors.
When stretched, the coupling becomes weaker and it is natural to
expect that atomic orbitals of the central atom give rise to the
resonance~\cite{Torres}.
%
Among the three contacts (Na, Pb, Al) on which we are going to
concentrate, the resonance formation is already demonstrated for
Na and Pb contacts~\cite{Cuevas_stretching,Yeyati,Sim,Hirose}. For
Al contacts, we present in this paper {\it ab initio} calculations
that confirm the resonance formation.

%
%
%
%

{\em Transport channels.--}
The number of transport channels depends on
the number of valence electrons~\cite{Scheer,Cuevas}; while
monovalent contacts have only one transport channel, polyvalent
contacts have multiple channels.
%
%
We begin our analysis by identifying ``eigen''-transport channels,
which are ``eigenvectors" of a proper transmission matrix and
decoupled from each other in this sense. The identification is
straightforward for a contact with a symmetric shape;
recent experiments~\cite{Rodrigues} showed that contacts tend to align
their axes with high symmetry lattice directions so that contacts
have a symmetric shape.
As an illustration, we describe the identification
of 3 channels~\cite{Cuevas} for a (111) Al contact with the $2\pi/3$
rotation symmetry and the inversion
symmetry~(Fig.~\ref{Alstructure}). The rotation symmetry with
respect to the $z$-axis (the horizontal axis in
Fig.~\ref{Alstructure}) makes the angular momentum $m$ $(=0,\pm
1)$ under the $2\pi/3$ rotation a good quantum number. Then a
traveling wave packet with a given $m$ can be scattered into
states with the same $m$ only and thus the three eigen-transport
channels of Al contacts can be characterized by $m$.
%
%
There is a natural link between the channels and the orbitals of
the central Al atom; $m=0$ channel couples with $3s$ and $3p_z$
orbitals, while $m=\pm1$ channels with $3p_x \pm i 3p_y$. Here the
factor $\pm i$ implies that the two channels $m=\pm 1$ are
mutually related by the time reversal and thus degenerate.
%

%
%
%


{\em Friedel sum rule.--} The conductance of a contact is given by
$G=G_0\sum_m T_m(E_F)$, where $T_m(E_F)$ is the transmission
probability of the channel $m$ at the Fermi energy $E_F$. In order
to examine the connection between $T_m$ and microscopic features,
it is useful to use the Friedel sum rule~\cite{Datta,Lee}.
%
%
%
For a contact with the inversion symmetry, which makes the parity
under the inversion a good quantum number, the Friedel sum
rule~\cite{Datta} results in
%
\begin{equation}
T_m(E_F)=\sin^2 \left({\pi \over 2}\Delta N_m\right),
\label{eigentran}
\end{equation}
where $\Delta N_m \equiv N_{m,e} - N_{m,o}$,
$N_{m,e(o)}=2\int^{E_F} dE \, \rho_{m,e(o)}(E)$, and
$\rho_{m,e/o}$ is the density of states (DOS) with even/odd parity
in the channel $m$ (spin degeneracy assumed).
%
%

{\em Resonant transport and charge neutrality.--} We then examine
how $\Delta N_m$'s are affected by microscopic features.
%
For the resonant transport,
$\Delta N_m$ is governed by resonances, $\Delta
N_m=2(R_{m,e}-R_{m,o})$, where $R_{m,e(o)}$ is the number of even
(odd) parity resonances in the channel $m$ below $E_F$ (2 from
spin degeneracy). Partially filled resonances (near $E_F$) give
fractional contribution to $R_{m,e/o}$.
%
%
%

Another important relation comes from the charge neutrality, which
is an excellent approximation in metallic
systems~\cite{Yeyati,Sim}. The charge neutrality near the central
atom results in $2\sum_m (R_{m,e}+R_{m,o})=N^\prime$, where
$N^\prime$ is the number of valence electrons in the central
atom~\cite{apex_resonance}.

To make a further progress, we first note that $T_m$ is periodic
in $R_{m,e/o}$ with the period 1 [Eq.~(\ref{eigentran})].
We thus ignore the integer part and focus on the fractional
part of $R_{m,e/o}$ arising from the partially filled resonance.
For contacts under consideration, moreover,
central atoms possess only a few atomic orbitals and as a result,
there is at best one partially filled resonance in each channel.
%
%
This is verified by our {\it ab initio} calculation given below
(Al contact) and by Refs.~\cite{Cuevas_stretching,Sim} (Pb and Na
contacts, respectively). Then for each $m$, either $R_{m,e}=0$ or
$R_{m,o}=0$, and the new equality $|\Delta
N_m|=2(R_{m,e}+R_{m,o})$ holds.
%
%
The charge neutrality condition then becomes
\begin{equation}
\sum_m |\Delta N_m|=N'_{\rm eff}, \label{neutrality}
\end{equation}
where $N'_{\rm eff}$ can differ from $N'$ by an integer multiple of 2
due to the neglect of the integer part of $R_{m,e/o}$.
%
%

{\em Mono. vs. polyvalent contacts.--} For monovalent contacts
such as Na~\cite{Sim}, where there is only one transport channel
and $N'=N'_{\rm eff}=1$, Eq.~(\ref{neutrality}) reduces to
$|\Delta N|=1$. According to Eq.~(\ref{eigentran}),
Eq.~(\ref{neutrality}) then implies $G=G_0$ regardless of other
details such as the degree of stretching, explaining the
quantization of $G$ in experiments.
%
%

In polyvalent contacts, on the other hand, there are multiple
channels and multiple $\Delta N_m$'s. Then the single
constraint~(\ref{neutrality}) alone cannot completely determine
$\Delta N_m$'s and $G$ needs not be quantized, explaining the
difference between monovalent and polyvalent contacts. This is one
of the main results of this paper.
%
%

As a concrete example of polyvalent contacts, we examine the
symmetric Al contact. When its central atom is weakly coupled to
the electrodes, its atomic orbitals $3s, 3p_x, 3p_y, 3p_z$ give
rise to one (almost) fully occupied resonance in the $m=0$ channel
(mostly $s$ character) and three partially filled resonances, one
in each channel. This occupation of resonances can be easily
inferred from the knowledge of an isolated Al atom. When the two
electrons in the fully occupied resonance are ignored, $N'_{\rm
eff}$ is $N'-2=1$ and from Eq.~(\ref{neutrality}), we have
\begin{equation}
|\Delta N_{0}|+2 |\Delta N_{\pm1}|=1, 
\label{Al_constraintL=1}
\end{equation}
where the degeneracy of the $m=\pm 1$ channels is used. From
Eq.~(\ref{eigentran}), the conductance then reads
\begin{eqnarray}
\frac{G}{G_0} =\sum_m T_m= \left[\cos\left(\pi \Delta
N_{\pm1}\right) - \frac{1}{2}\right]^2
 + \frac{3}{4}\; .
\label{L=1Al}
\end{eqnarray}
Thus $G$ {\it varies} as $\Delta N_{\pm 1}$ changes during the
stretching.

{\em Orbital shapes.--} The value of $\Delta N_{\pm 1}$ is
influenced by the shapes of $s$ and $p$ orbitals, which make the
$pp\pi$ coupling (responsible for the partially filled $m=\pm 1$
resonances) weaker than the $pp\sigma$ or $sp\sigma$ coupling
(responsible for the partially filled $m=0$ resonance).
Consequently $|\Delta N_{\pm 1}|< |\Delta N_0|$, and from
Eq.~(\ref{Al_constraintL=1}), we find $|\Delta N_{\pm 1}|<1/3$.
The stretching magnifies the effect of the orbital shape
difference~\cite{Cuevas_stretching,Handbook} and thus decreases 
$|\Delta N_{\pm 1}|$
further below $1/3$. Then according to Eq.~(\ref{L=1Al}), $G$ {\em
increases} monotonically during the stretching, in agreement with
experiments~\cite{Scheer,SanchezPortal,Cuevas_stretching}. 
Our explanation for Al contacts is in qualitative agreement with 
Ref.~\cite{Cuevas_stretching} but differs from Ref.~\cite{Kobayashi},
where the displacement of the central atom from the $z$-axis is important.
Equation~(\ref{L=1Al}) also predicts that $G$
has an upper bound of $G_0$ and a lower bound of 0.75 $G_0$, in
reasonable agreement with the histogram
data~\cite{Yanson_comment,bounds_comment}, 
and that $T_0$ ($sp$
contribution) increases during the stretching, while $T_{1}$ and
$T_{-1}$ decrease.
Equation~(\ref{L=1Al}) and its implications are the second main
result of this paper.



{\em Pb contacts.--} Next we apply this approach to Pb contacts,
where 6$p$ orbitals mediate the transport. In an isolated Pb atom,
the spin-orbit coupling splits them into $p_{1/2}$ and $p_{3/2}$
with the former lower in energy by about 0.9 eV \cite{Wurde}. In a
Pb contact with the $2\pi/3$ rotational symmetry, its channels
then can be
%
%
classified as $(j,m_j) = (1/2,
\pm 1/2)$, $(3/2,\pm 3/2)$, and $(3/2, \pm 1/2)$, where $j$ is the
total angular momentum and $m_j$ is its component along the
$z$-axis. Since each pair of channels $(j,\pm m_j)$ are
degenerate, the $\pm$ sign can be regarded as a fictitious spin
variable for the channels. Then we have three transport channels,
each with the ``spin'' degeneracy two, and obtain
$T_{(j,|m_j|)}=\sin^2(\pi\Delta N_{(j,|m_j|)}/2)$, where $\Delta
N_{(j,|m_j|)}=2\int^{E_F} dE \, [ \rho_{(j,\pm
m_j),e}(E)-\rho_{(j,\pm m_j),o}(E)]$.

Since $N'_{\rm eff}=2$ for a Pb atom (2 electrons in the $6s$
orbital are ignored), one obtains, instead of
Eq.~(\ref{Al_constraintL=1}),
\begin{equation}
\sum_{(j,|m_j|)} |\Delta N_{(j,|m_j|)}| = 2.
\label{Pb_constraintL=1}
\end{equation}
The role of the constraint~(\ref{Pb_constraintL=1}) becomes clear
in the large $d$ limit, where the orbital coupling to electrodes
is weak and $|\Delta N_{(3/2,3/2)}|\simeq |\Delta
N_{(3/2,1/2)}|<|\Delta N_{(1/2,1/2)}|$.
Equation~(\ref{Pb_constraintL=1}) then predicts
$T_{(3/2,3/2)}\simeq T_{(3/2,1/2)}<T_{(1/2,1/2)}$, which agrees
with a channel-resolved measurement~\cite{Cuevas_stretching}. Note
that except for the difference in $N'$ (or $N'_{\rm eff})$, Pb and
Al contacts are very similar. The difference in $N'$ is however
crucial; from Eq.~(\ref{Pb_constraintL=1}), one obtains
\begin{equation}
{G \over G_0} \simeq {9 \over 4}-\left[ {1\over 2}+\cos \left( \pi
\Delta N_{(3/2,3/2)} \right) \right]^2, \label{L=1Pb}
\end{equation}
which differs from Eq.~(\ref{L=1Al}).
We next consider the behavior of $\Delta N_{(3/2,3/2)}$.
Since the $(3/2,3/2)$ channel is least favored by both orbital
and spin-orbit couplings to electrodes, $|\Delta N_{(3/2,3/2)}|$
should be smaller than $2/3$
[Eq.~(\ref{Pb_constraintL=1})]
and decay during the stretching. From Eq.~(\ref{L=1Pb}), $G/G_0$ then
decays monotonically during the stretching with the upper (lower)
bound of $9/4$ ($0$), which is in reasonable agreement with
experiments~\cite{Cuevas_stretching}. For realistic values of $d$,
on the other hand, the orbital coupling will lift the degeneracy
$|\Delta N_{(3/2,3/2)}|=|\Delta N_{(3/2,1/2)}|$.
However this effect turns out to be rather
minor: A recent calculation~\cite{Cuevas_stretching} suggests that
already at $d=1.2 \, d_{0, {\rm Pb}}$, where $d_{0, {\rm Pb}}$ is
the bond length of bulk Pb, $|\Delta N_{(3/2,3/2)}|$ and $|\Delta
N_{(3/2,1/2)}|$ ($<1$) are significantly smaller than $|\Delta
N_{(1/2,1/2)}|$ ($>1$). Then the predictions of Eq.~(\ref{L=1Pb})
should remain qualitatively valid at least for $d>1.2 \, d_{0,{\rm
Pb}}$. We also note that in the range of $d$, where $|\Delta
N_{(1/2,1/2)}|$\- $>1$, not only $G$ but also each individual
$T_{(j,|m_j|)}$ {\em decreases} during the stretching due to
Eq.~(\ref{Pb_constraintL=1}). This part on Pb contacts constitutes
the third main result of this paper.


\begin{figure}
\includegraphics[width=0.40\textwidth]{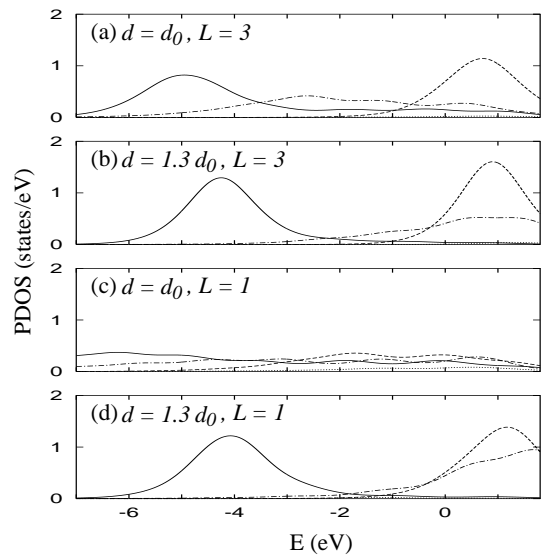}
\caption{ Projected DOSs (PDOS) onto the central atom of Al
contacts with different $d$ and $L$. Solid and dash-dotted (dotted
and dashed) lines represent the PDOSs for the $m=0$ ($m=\pm 1$)
channel with even and odd parity, respectively. $E_F$ is aligned
at 0 eV and the PDOS for $m=0$ with odd parity is magnified by
factor 2 for clarity. } \label{PDOS}
\end{figure}

{\em Resonance in Al contacts: {\it Ab initio} calculation.--}
While the resonance formation is already verified for
Na~\cite{Yeyati,Sim,Hirose} and Pb~\cite{Cuevas_stretching} contacts,
it remains rather unclear for Al contacts~\cite{Hopping}.
%
%
We thus perform first-principles calculation for an Al contact
with the geometry in Fig.~\ref{Alstructure} ($L=3$).
Pseudopotentials \cite{pseudo} and real-space multigrid method
\cite{Jin} within the local-density-functional approximation are
used. Each electrode is modeled by a fcc cluster of $M$ ($= 69$)
Al atoms. Small cluster calculations are useful to test the
resonance formation although the predicted resonance positions may
be wrong due to the finite size effect (at best $\lesssim 1$ eV in
our estimation). A supercell geometry containing the whole system
is used and electrodes in neighboring supercells are separated by
more than 10.58 \AA, so that inter-supercell interactions are
negligible. Using a grid spacing of 0.44 \AA, the total energy is
converged to within $10^{-7}$ Ry. For the $M=69$ cluster, single
particle level spacing $\delta \epsilon$ is estimated at $\sim
0.07$ eV near $E_F$. To extract properties of infinite metallic
electrodes from the finite cluster calculation, the Fermi-Dirac
level broadening $E_T$ should be larger than $\delta \epsilon$.
Using $E_T = 0.4$ eV ($> 3 \delta \epsilon$), we obtain the
projected DOSs (PDOS) of different $m$ states onto a region around
the central atom (a sphere with the radius of $0.56\, d_0$) and
decompose them into even and odd parity components with respect to
the inversion symmetry. Here, $d_0$ ($ = 2.86$ \AA) is the bond
length of Al fcc metals.

The resonance assumption is tested for $d=d_0$ and $1.3 \,
d_0$~(Fig.~\ref{PDOS}). For $d=1.3 \, d_0$, the stretching
distance is $2(d-d_0)=0.6 \, d_0$, which is smaller than the
experimental stretching distance of $0.7\, d_0$ to $1.7\, d_0$
before the contact
breaking~\cite{Scheer,SanchezPortal,Cuevas_stretching}.
Figure~\ref{PDOS}(b) shows the PDOSs for $d=1.3\, d_0$. In the
$m=0$ channel, two well-defined resonances develop, one with even
parity (almost fully filled) and the other with odd parity.
In the $m=\pm 1$ channel, one odd parity resonance develops. Thus
for $d=1.3 \, d_0$, the resonance assumption holds for all
channels. When $d$ is reduced to $d_0$ [Fig.~\ref{PDOS}(a)],
channels respond differently due to the difference in the coupling
strengths between the $\sigma$ and $\pi$ couplings. Whereas the
resonance in the $m=\pm 1$ channel remains well-behaved,
resonances in the $m=0$ channel become broader and make
nonnegligible overlap at $E_F$. Thus not all channels are in the
clear resonant regime. Recalling that the resonance widths in
Fig.~\ref{PDOS} can be overestimated by $E_T$, we may safely
conclude that Al contacts enter the clear resonant regime
somewhere between $d_0$ and $1.3 \, d_0$. For $d_0 \lesssim d
\lesssim 1.3d_0$, the correction to $G$ from the $m=0$ channel
being not in the clear resonant regime is $-\sin (\pi\eta) \sin
[\pi(\eta+2 \Delta N_{\pm1})]$, where $\eta\equiv 2-2R_{m=0,e}$.
Thus for $\eta\ll 1$, the correction is not significant and the
general trend of rising $G$ during the stretching is not altered.

The PDOSs are calculated for the $L=1$ geometry model of contacts
as well, where the two electrodes share a common apex atom. The
PDOSs for $d=d_0$ [Fig.~\ref{PDOS}(c)] are essentially
structureless, implying that the sharp tip geometry alone does not
induce resonances at Al apex atoms, in contrast to the result for
the Na (111) tip geometry~\cite{Yeyati,Sim}. But the resonance
assumption holds for $d=1.3 \, d_0$ [Fig.~\ref{PDOS}(d)] and a
conclusion similar to that for the $L=3$ geometry can be made.

{\em Discussion and conclusion.--}
%
%
%
We first note that although contacts with a symmetric shape are
used for analysis, deviations from the symmetric geometry
such as disorder in atomic positions do not affect
the result qualitatively, as demonstrated explicitly 
in Refs.~\cite{Yeyati,Cuevas}.
Secondly, our analysis can be used to study effects of $L$
variation as well.
For monovalent contacts, for example, the resonance analysis
predicts different conductance behavior for even and odd $L$
due to the difference in $N'_{\rm eff}$.
This prediction is consistent with the oscillation of $G$ with $L$
in Na contacts (first-principles calculations~\cite{Sim,Hirose}).
In Au, where $L$ is known~\cite{Bahn} to vary during the stretching,
a similar oscillation was indeed observed~\cite{Rubio-Bollinger,Smit},
although its amplitude is smaller than the predicted value for Na 
probably due to the stronger coupling in Au~\cite{Hopping}.
%
%
Thirdly, our analysis can be also extended to contacts made of $sd$
metal, which have five channels and thus could be more complex. It
was recently reported~\cite{Nielsen} that a one-atom Pt contact
has effectively three open channels ($sd_{z^2}$, $d_{yz}$, and
$d_{zx}$ characters) at the Fermi level under zero bias. Here, one
can find $N'_{\rm eff} = 4$ since 6 among 10 valence electrons of
a Pt atom participate in the almost fully filled $sd_{z^2}$,
$d_{xy}$, and $d_{x^2-y^2}$ states~\cite{Nielsen}. Then, our
resonance analysis leads to $G/G_0 \simeq 9/4 - (1/2 + \cos\pi
\Delta N_{d_{yz}/d_{zx}})^2$ and 
$|\Delta N_{sd_{z^2}}| + 2 |\Delta N_{d_{yz}/d_{zx}}| = 4$.
Because $|\Delta N_{d_{yz}/d_{zx}}|$ is larger than 4/3 and
increases (at best up to 2) during the
stretching~\cite{Pt_contact}, $G$ decreases during the stretching
with the upper bound of $\sim 9/4 G_0$, in good agreement with
Ref.~\cite{Nielsen}.
Fourthly, we remark that our approach may be also useful for a
wider class of systems, for example, symmetric molecules coupled
to electrodes~\cite{Reichert}. For this purpose, the charge
neutrality condition~(\ref{neutrality}) should be generalized to
take into account possible electron transfer between molecules and
electrodes arising from their electron affinity difference.
%
%
%

To conclude, under the single assumption of resonant transport, it
is demonstrated that qualitatively different conductance behavior
in various stretched contacts can be explained in a unified way in
terms of the number of electrons participating in partially filled
resonance states, orbital shapes, and charge neutrality.

HSS, DHK, and KJC were supported by KISTI and QSRC at Dongkuk
University, and HWL was supported by SKORE-A, eSSC, 
Nano Research and Development Programs, and POSTECH BSRI 
research fund.

\bibliographystyle{apsrev}

\begin{thebibliography}{99}
\bibitem{difference}
In Ref.~\cite{Cuevas_stretching}, tight-binding calculation has been
performed under the constraint of the local charge neutrality.
A stretched contact is modeled as a contact with enlarged atomic
spacing at the contact region and 
it has been found that the result 
is in qualitative agreement with the
experimental observation for Al contacts when a sufficiently large number
of atomic layers from the central atom of the contact  
has been taken into account~\cite{Yeyati}.  
In Ref.~\cite{Kobayashi}, on the other hand, the first-principles
calculation has been performed. Due to the calculational cost, only
one atomic layer has been taken into account.
From this calculation, it has been reported that 
the enlarged atomic spacing alone cannot explain the experimental
finding and the behavior of the increasing conductance during the
stretching is achieved only when 
the atomic arrangement at the
contact region forms a ``bent'' wire.
\bibitem{L=1vsL=3} It is not yet clear whether monatomic
contacts in experiments correspond to the $L=3$ geometry in
Fig.~\ref{Alstructure} or an $L=1$ one, where two leads share a
common apex atom. Our results for both cases are the qualitatively
same.
\bibitem{apex_resonance}
The charge neutrality condition may acquire a correction when the
two apex atoms of the electrodes give rise to resonances. Such
apex resonances are however irrelevant for atomic contacts under
consideration.
\bibitem{bounds_comment}
These bounds can be weakly violated due to various details ignored
in this idealized analysis.
\bibitem{Pt_contact}
The partially filled $sd_{z^2}$ resonance has antibonding coupling
to electrodes, thus has higher energy than those of $d_{yz}$ and
$d_{zx}$. The details will be published elsewhere.
\end{thebibliography}



\end{document}